\documentclass[epsf,12pt]{article}

\begin{document}

\title{ Transient of the kinetic spherical model
between two temperatures
\thanks{Supported by the National Natural Science
Foundation of China(Grant No. A050105) and the Advance Research
Center of Zhongshan University} }
\author{\normalsize \sf Chunshan He and
Zhibing Li \footnote{Corresponding author, email:stslzb@zsu.edu.cn}   \\
\normalsize Department of Physics,  Zhongshan University,
Guangzhou 510275, China  }
\date{March, 2003}

\maketitle
\vskip 1.pc

\vskip 1.5pc
\begin{abstract}
We solve the dynamic equation for the kinetic spherical model that
initially is in an arbitrary equilibrium state and then is left to
evolve in a heat-bath with another temperature. Flows of the
Renormalizational group are determined.



Shortened version of the title: Critical dynamics of the kinetic
spherical model
\end{abstract}
\vskip 6pt {\sf }

{

\newpage
\section*{1. Introduction}

In recent years, the universal scaling in non-equilibrium states
have attracted much
attention\cite{zheng98,sari2000,qing2001,scop2002,biva2002}. The
phase-ordering process (POP) \cite{s3} and the short-time critical
dyanmics (SCD) \cite{janss} are two fruitful examples. In the POP,
the system initially at a very high temperature then is quenched
to a heat-bath of very low temperature. In the SCD, the heat-bath
has the critical temperature of the system.

In both POP and SCD, the initial correlation length is assumed to
be zero. For finite initial correlation, the scaling invariance of
the initial state is broken. One would expect a crossover
phenomenon that is usually quite difficult to study either by
theoretic methods or by numerical simulation. In order to gain an
insight into this phenomenon, in the present paper we investigate
the kinetic spherical model (KSM) with initial correlation.

The short-time dynamics of the kinetic spherical model with zero
initial correlation length has been studied in
\cite{zheng96,chen00,hw93,hw95,coni94,kiss93,huma91,bray90,bray92}.
In the latest publication \cite{chen00} the effect of non-zero
initial order was emphasized.

The system to be considered in the present paper is initially at
an arbitrary equilibrium state. Then it is suddenly put into a
heat-bath of another temperature. The external field is also
removed instantly. The system is assumed to evolve following the
Langevin equation. We are interested in the non-equilibrium
transient state in the following time. We will concentrate on the
time-dependent order parameter $m(t)$. To include the non-zero
correlation, one must integrate the non-trivial properties of the
initial equilibrium state. Though most properties of the spherical
model can be found in literatures, e.g., the famous book by Baxter
\cite{baxt82}, it is still quite tricky to connect the equilibrium
initial state with the non-equilibrium dynamics. We find that
besides the subtraction of mass that is known in the zero initial
correlation, one need an extra renormalization of the dynamic
equation.

\section*{2. The Model}

\par
   The Hamiltonian of the spherical model is
   \begin{equation}
   \label{ee1}
   H={\frac{\alpha}{2}}\sum\limits_{i} S ^{2}_{i}-
   \beta J \sum\limits_{<ij>} S_{i}S_{j}-\beta
   \sum\limits_{i}h_{i}S_{i}
   \end{equation}
with the constraint
   \begin{equation}
   \label{ee2}
   {\sum\limits_{i} S ^{2}_{i}}=N
   \end{equation}
where $<ij>$ are bonds of a 3-dimensional regular lattice, $N$ is
the total number of spins; $\beta={\frac {1}{k_{B} T}}$. In the
dynamic process, $\alpha$ is a time-dependent Lagrange multiplier
corresponding to the constraint.

The fluctuations of spins are defined to be
  $ \widetilde S _{i}=S_{i}-<S_{i}>$.
In the momentum space, one has
\begin{equation}
   \label{ee3}
\widetilde S (p,t)={\frac{1}{\sqrt{N}}}\sum\limits_{i} \widetilde
S_{i} e^{i {\bf p}\cdot {\bf r_{i}}} \end{equation} with $r_{i}$
the position vector of site $i$. Define
\begin{equation}
   \label{ee4}
\Omega(p)=1-{\frac{1}{3}}(cos(p_{1})+cos(p_{2})+cos(p_{3}))
\end{equation}
and a function
    \begin{equation}
    \label{ee5}
    w(x)={\frac{1}{N}}\sum\limits_{p} {\frac{1}{x+
    \Omega (p)}}
    \end{equation}
The correlation length is inversely proportional to the square
root of $z_{0}$ that is the solution of the equation
\begin{equation}
\label{ee6}
w(2z_{0})=\beta_{0}J-\frac{\beta_{0}h^{2}}{4Jz^{2}_{0}}
\end{equation}
for the initial temperature $T_{0}$ ($\beta_{0}=1/k_{B}T_{0}$) and
the homogenous field $h$ . The initial magnetization is given by
\begin{equation}
\label{ee7} m_{0}=\frac{h}{2Jz_{0}}
\end{equation}
The equation of equilibrium state turns out to be
\begin{equation}
\label{ee8} w(\frac{h}{Jm_{0}})=\beta_{0}J(1-m^{2}_{0})
\end{equation}

The initial correlation function is
\begin{equation}
\label{ee9} C_{0}(p,p^{\prime})=<\widetilde S (p,0)\widetilde S
(p^{\prime},0)>=\frac{1}{\beta_{0}J(2z_{0}+\Omega(p))}\delta_{p,p^{\prime}}
\end{equation}
It is easy to see that
\begin{equation}
\label{ee10} w(2z_{0})=\frac{\beta_{0}J}{N}\sum\limits_{p}
C_{0}(p,p)
\end{equation}

\par
The Langevin equation in the momentum space for this constrained
spin system is \cite{chen00}
   \begin{equation}
   \label{ee11}
   {\frac{\partial \widetilde S (p,t)}{\partial t}}=-\lambda
   (\tau (t)+\beta J \Omega (p))\widetilde S (p,t)+\eta (p,t)
   \end{equation}
where $T$ ($\beta=1/k_{B}T$) is the temperature of the heat-bath.
The consistency condition gives
   \begin{eqnarray}
   \label{ee12}
   \tau (t) & = & \tau_{sub}+\beta J\left [ m^{2}
   (t)-1\right ] +
   {\frac{\beta J}{N}}
   \sum\limits_{p} (1-\Omega(p))< \widetilde S (-p,t)
   \widetilde S(p,t)> \nonumber \\
   & + &
   {\frac{1}{\lambda N}}
   \sum\limits_{p}< \widetilde S (-p,t)\eta (p,t)>
   \end{eqnarray}
where the first term comes from the mass subtraction which
guarantees $\tau_{c}(\infty)=0$ at the
critical point. Recalling that the
equilibrium correlation
has a zero pole at the critical temperature,
one has
 $$< \widetilde S (-p,\infty)\widetilde S(p,\infty)>\vert_{\beta_{c}}
\sim {1 \over \Omega(p)}$$ It defines the critical temperature
$\beta_{c} J = w(0)$ \cite{baxt82}. Substituting it into
(\ref{ee12}) at the critical temperature, one can find
$\tau_{sub}=1$ for the Ito prescription with which the last term
of (\ref{ee12}) is zero due to causality. If the Stratonovich
prescription is used, the last term of (\ref{ee12}) is $1$ and
$\tau_{sub}=0$. Through the paper, Ito prescription will be used.

\par
By solving (\ref{ee11}), it is not difficult to obtain the
response propagator
  \begin{equation}
  \label{ee13}
  G_{p}(t,t^{\prime})={\frac {1}{2\lambda}}
  < \widetilde S (-p,t)\eta (p,t^{\prime})>
  =\Theta (t-t^{\prime})e^{-\lambda \beta J \Omega(p)
  (t-t^{\prime})-\lambda\int\limits_{t^{\prime}
  }^{t}dt^{\prime\prime}
  \tau (t^{\prime\prime})}
  \end{equation}
with the Heaviside step function $ \Theta (t-t^{\prime})= 1 $ for
$t>t^{\prime}$, otherwise $\Theta (t-t^{\prime})=0$; and the full
correlation function (correlation function including the initial
correlation)
  \begin{eqnarray}
  \label{ee14}
  \widetilde C_{p}(t,t') & = &
  < \widetilde S (p,t)\widetilde S (-p,t^{\prime})> \nonumber \\
  & = &
  C_{0}(p,p)
  G_{p}(t,0)G_{-p}(t^{\prime},0)+C_{p}(t,t^{\prime})
  \end{eqnarray}
with the correlation function
  \begin{equation}
  \label{ee15}
  C_{p}(t,t^{\prime})=2\lambda\int\limits_{0}^{\infty}
  dt^{\prime\prime}G_{p}(t,t^{\prime\prime})
  G_{p}(t^{\prime},t^{\prime\prime})
  \end{equation}

\section*{3. Laplace transformation}
\par
Introducing $ f(t)=m^{-2} (t)$ with the time-dependent
magnetization $m(t)=\frac{1}{N}<\sum\limits_{i}S_{i}>$, one can
convert the dynamic equation into a linear integrodifferential
equation for $f(t)$
    \begin{equation}
    \label{ee16}
    {\frac {\partial f(t)} {\partial t}}=2\lambda\beta J
    -2\lambda( \beta J -1)f(t)+{\frac{2\lambda\beta J}{N}}
    \sum\limits_{p} (1-\Omega(p)) \widetilde C_{p}(t,t)f(t)
    \end{equation}
For convenient, define $\gamma=(2\lambda \beta J)^{-1}$. By
Laplace transformation
    $$F(q)=\int\limits_{0}^{\infty}  dt f(t) e^{-qt}  $$
Equation (\ref{ee16}) is transformed to
        \begin{equation}
        \label{ee17}
   F(q)=\frac{
    {\frac{1}{q}} +\frac{\gamma}{m^{2}_{0}}\frac{1}{N}\sum\limits_{p}\frac{C_{0}(p,p)}{\gamma q
    +\Omega(p)}}{1 -\frac{1}{\beta J}w(\gamma q)}
     \end{equation}
Substituting $C_{0}(p,p)$ by equation (\ref{ee9}), one has
        \begin{eqnarray}
        \label{ee18}
   F(q)&=&\frac{
    {\frac{1}{q}} +\frac{\gamma}{\beta_{0}
    J m^{2}_{0}}\frac{1}{N}\sum\limits_{p}\frac{1}{(\gamma q+\Omega(p))
    (2z_{0}+\Omega(p))}}{1 -\frac{1}{\beta J}w(\gamma)} \nonumber \\
    &=&\frac{
    {\frac{1}{q}} +\frac{\gamma}{\beta_{0}
    J m^{2}_{0}}\frac{1}{2z_{0}-\gamma q}\left[w(\gamma q)-w(2z_{0})\right]}
    {1 -\frac{1}{\beta J}w(\gamma q)}
     \end{eqnarray}
The last square bracket in the numerator can be written as
\begin{equation}
\label{ee19}
 w(\gamma q)-w(2z_{0})=\left[ \beta J-w(2z_{0})\right ]-\Lambda
 \left[ \beta J-w(\gamma q) \right]
\end{equation}
Where the constant $\Lambda=1$. In fact, we will see that
$\Lambda$ plays as a renormalization multiplier that should be
determined self-consistently. The reason will be clear soon. For
the infinite system the sum in (\ref{ee5}) is replaced by an
integral. In the continuum-limit one has
   \begin{equation}
   \label{ee20}
   w(\gamma q)=w(0)-D(\gamma q)^{1/2}
   \end{equation}
where the constant $D=({\frac{9}{2\pi^{2}}})^{3/2}$. In this
expansion of $w(x)$, the spatial and temporal microscopic detail
is lost. However, this microscopic detail would have macroscopic
effects through equation (\ref{ee19}) since it associates with the
factor
$$ \frac{1}{2z_{0}-\gamma q}$$
which in Laplace reversion is a factor increasing versus time
exponentially. Therefore, a renormalization multiplier $\Lambda$
is needed to rescue the error introduced by equation (\ref{ee19}).

Recalling $q$ has the inverse unit of time, one easily finds two
characteristic time-scales
\begin{eqnarray}\label{ee21}
t_{h}&=&\frac{\gamma}{2 z_{0}}=\frac{m_{0}}{2\lambda \beta h} \nonumber \\
t_{\beta}&=&\left[ \frac{2\lambda \gamma^{3/2} D}{\mu}\right ]^{2}
\end{eqnarray}
where $\mu=1-T/T_{c}$. By use of (\ref{ee19}), $F(q)$ is written
as
\begin{eqnarray}
\label{ee22}
 F(q)&=&\frac{1}{\mu q(1+\sqrt{t_{\beta}q})}-\frac{\Lambda \beta
t_{h}}{\beta_{0} m^{2}_{0}}\frac{1}{1-t_{h}q} \nonumber \\
&+& \frac{A}{\mu}\frac{1}{1-t_{h}q}\frac{1}{1+\sqrt{t_{\beta}q}}
\end{eqnarray}
where
\begin{equation}\label{ee23}
A=\frac{\beta (1-\frac{1}{\beta
J}w(2z_{0}))}{\beta_{0}m^{2}_{0}}={\frac{\beta
J-w(\frac{h}{Jm_{0}})}{w(\frac{h}{Jm_{0}})}}{\frac{1-m^{2}_{0}}{m^{2}_{0}}}
\end{equation}

When the heat-bath is at the critical temperature, $F(q)$ is
\begin{eqnarray}
\label{ee24}
 F(q)&=&B q^{-3/2} - \frac{\Lambda \beta_{c}
t_{h}}{\beta_{0}m^{2}_{0}}\frac{1}{1-t_{h}q} + \frac{A
B}{(1-t_{h}q)q^{1/2}}
\end{eqnarray}
where $B=\frac{(2\lambda\beta_{c}J)^{3/2}}{2\lambda D}$.

\section*{ 4. Laplace reversion}

Let us first consider the case of $T < T_{c}$, i.e., $\mu > 0$.

A direct Laplace reversion to (\ref{ee22}) gives
   \begin{eqnarray}
   \label{ee25}
    m(t)&=& \sqrt{\mu} \{
    1+ {A t_{h}\over t_{\beta}- t_{h}}\sqrt{{t_{\beta}\over
    t_{h}}}e^{t/t_{h}}
    erfc(\sqrt{{t \over t_{h}}}) \nonumber  \\
    &-& \left [ 1+{A t_{h} \over {t_{\beta}- t_{h}}} \right ]
    e^{t/t_{\beta}} erfc(\sqrt{{\frac{t}{t_{\beta}}}}) \nonumber \\
    &+&\left [{\frac{\beta
    \mu}{\beta_{0}m^{2}_{0}}}\Lambda-{\frac{A
    \sqrt{t_{h}}}{\sqrt{t_{\beta}}-\sqrt{t_{h}}}}
    \right ] e^{t/t_{h}} \}^{-1/2}
    \end{eqnarray}
where
\begin{equation}
\label{ee26} erfc({x})={\frac {2}{\sqrt{\pi}}}\int\limits_{x}^{
\infty} e^{-\tau^{2}} d\tau
\end{equation}
is the complementary error function. For large $x$, it has the
asymptotic expansion
\begin{equation}\label{ee27}
erfc(x)=\frac{e^{-x}}{\sqrt{\pi}}\left(x^{-1/2}-\frac{1}{2}x^{-3/2}+\frac{3}{4}x^{-5/2}+\cdots
\right)
\end{equation}
In equation (\ref{ee25}), the last term is not welcome since it
leads to a fault exponential decay of magnetization. It can be
killed by choosing the renormalization multiplier as
\begin{equation}\label{ee28}
\Lambda={\frac{A
    \sqrt{t_{h}}}{\sqrt{t_{\beta}}-\sqrt{t_{h}}}}\frac{\beta_{0}m^{2}_{0}}{\beta
    \mu}
\end{equation}
One can show that $\Lambda = 1$ in the limit $z_{0}\rightarrow 0$.
The final result is
    \begin{eqnarray}
      \label{ee29}
    m(t)&=& \sqrt{\mu} \{
    1+ {A t_{h}\over t_{\beta}- t_{h}}\sqrt{{t_{\beta}\over
    t_{h}}}e^{t/t_{h}}
    erfc(\sqrt{{t \over t_{h}}}) \nonumber \\
    &-& \left [ 1+{A t_{h} \over t_{\beta}- t_{h}}\right ]
    e^{t/t_{\beta}} erfc(\sqrt{{\frac{t}{t_{\beta}}}})\}^{-1/2}
    \end{eqnarray}

The case of $T>T_{c}$ can be attained by similar method
   \begin{eqnarray}
      \label{ee30}
    m(t)&=&\sqrt{|\mu|}
    \{ -1+{\frac{A t_{h}}{t_{\beta}-{t_{h}}}} \sqrt{{\frac{t_{\beta}}{t_{h}}}}e^{t/t_{h}}
    erfc(\sqrt{{\frac{t}{t_{h}}}}) \nonumber \\
    &+& \left [1+{\frac{A t_{h}}{t_{\beta}-t_{h}}}\right ]
    e^{t/t_{\beta}}(2-erfc(\sqrt{\frac{t}{t_{\beta}}}))
     \}^{-1/2}
    \end{eqnarray}

One can attain the critical evolution of magnetization from the
Laplace reversion of (\ref{ee24}),
\begin{equation}\label{ee31}
m(t)=\left [ \frac{B^{2}t_{h}}{\pi}\right ]^{-1/4} \left [ 2( {t
\over t_{h}})^{1/2} + A({t_{h} \over
t})^{1/2}g(\frac{t_{h}}{t})\right ]^{-1/2}
\end{equation}
where
\begin{equation}\label{ee32}
g(x)=\int^{\infty}_{0}dy \frac{e^{-y}}{(1+xy)^{1/2}}
\end{equation}
It is clear that $g(0)=1$. The critical behavior also can be
attained by taking the limitation $T\rightarrow T_{c}$ in equation
(\ref{ee29}).

To recover the results of zero initial correlation of
\cite{chen00}, one only need to take the limitation of
$\beta_{0}\rightarrow \infty$, that is $z_{0}\rightarrow \infty$
(or $t_{h} \rightarrow 0$), in (\ref{ee29}), (\ref{ee30}) and
(\ref{ee31}), respectively. In this limit, $A=\frac{2
t_{i}}{t_{h}}$ with $t_{i}$ defined in \cite{chen00} and
$t_{\beta}=\pi t_{\mu}$.

\section*{ 5. Discussions and Conclusions}

In summary, we have studied the transient behavior of KSM that is
quenched from an arbitrary temperature into another. The formula
can describe ordering/disordering phenomena and the critical
dynamics. We find that the correct long-time behavior can be only
recovered after the subtraction of mass as well as the
renormalization of the dynamic equation.

From the exact magnetization obtained in the present paper, one
can find out the flows of renormalizational group of the bare
parameters $\beta$, $z_{0}$ and $m_{0}$(or $\beta_{0}$) under the
scale transformation. Changing the time scale by a factor $b^{z}$,
in order to keep the macroscopic quantity $m(t)$ unchange up to a
scaling factor, $t_{h}$ and $t_{\beta}$ must be transformed in the
same way,
\begin{equation}\label{ee33}
t_{h}(b) = b^{z} t_{h}(1), t_{\beta}(b) = b^{z} t_{\beta}(1)
\end{equation}
while $A$ must be an invariant,
\begin{equation}
A(b) = A(1)
\end{equation}
The relations of bare parameters and the scaling factor $b$ are
implicitly defined in the above equations. These are the so-called
characteristic functions for the dynamic crossover phenomena
\cite{zheng96, chen00, li02, he03}.


\vskip 1.pc

\begin{thebibliography}{99}
\bibitem{zheng98}B. Zheng, Int. J. Mod. Phys. B {\bf 12}, 1419(1998)
\bibitem{sari2000}Sarika Bhattacharyya, Biman Bagchi, Phys. Rev. E {\bf 61}, 3850(2000)
\bibitem{qing2001}Qing-Hu Chen, Meng-Bo Luo, Zheng-Kuan Jiao, Phys. Rev. B {\bf 64}, 212403(2001)
\bibitem{scop2002}T. Scopigno, {\it et al.}, Phys. Rev. Lett. { \bf 89}, 255506(2002)
\bibitem{biva2002}Bivash R. Dasgupta, {\it et al.}, Phys. Rev. E {\bf 65}, 051505(2002)
\bibitem{s3}A. J. Bray, Adv. Phys. {\bf43}, 357(1994)
\bibitem{janss}H. K. Janssen, in {\it From Phase Transition to Chaos},
edited by G. Gy\"orgyi, I. Kondor, L. Sasv\'ari, T. T\'el, Topics
in Mondern Statistical Physics (World Scientific, Singapore, 1992)
\bibitem{zheng96}B. Zheng, Phys. Rev. Lett. { \bf 77}, 679(1996)
\bibitem{chen00}Y. Chen, Shuohong Guo, Zhibing Li, Aijun Ye, Eur. Phys. J. B {\bf 15}, 97(2000)
\bibitem{hw93}H. W. Diehl, U. Ritschel, J. Stat. Phys. {\bf 73}, 1(1993)
\bibitem{hw95}U. Ritschel, H. W. Diehl, Phys. Rev. E {\bf 51}, 5392(1995)
\bibitem{coni94}A. Coniglio, P. Ruggiero, M. Zannetti, Phys. Rev. E {\bf50},
1046(1994)
\bibitem{kiss93}J. G. Kissner, A. J. Bray, J. Phys. A: Math. Gen. {\bf26}, 1571
(1993)
\bibitem{huma91}K. Humayun, A. J. Bray, J. Phys. A: Math. Gen. {\bf 24},
1915 (1991)
\bibitem{bray90}A. J. Bray, Phys. Rev. B {\bf 41}, 6724(1990)
\bibitem{bray92}A. J. Bray, J. G. Kissner, J. Phys. A: Math. Gen. {\bf 25},
31 (1992)
\bibitem{baxt82}R. J. Baxter, in {\it Exactly Solved Models in Statistical Mechanics} (Academic Press, New York,
1982)
\bibitem{li02}Z. B. Li, S. P. Seto, M. Y. Wu, H. Fang, C. S. He, Y. Chen, Phys. Rev. E, {\bf 65}, 057101 (2002)
\bibitem{he03}C. S. He, H. Fang, Z. B. Li, Science in China G, Vol. {\bf 46} No. {\bf 1}, 98 (2003)
\end{thebibliography}
\end{document}